\documentclass[showpacs,amsmath,amssymb,prb,aps,twocolumn]{revtex4}
\usepackage{graphicx}
\begin{document}
\title {Integral charge quasiparticles in a fractional quantum Hall liquid}
\author {Giovanni Vignale}
\email{vignaleg@missouri.edu}
\affiliation{Department of Physics and Astronomy,  University of Missouri,
Columbia, Missouri 65211, USA}
\date{\today}
\begin{abstract}
Starting from a collective description of the incompressible fractional quantum Hall liquid as an elastic medium that supports gapped neutral excitations  I show that the one-electron spectral function of this system exhibits a sharp peak at the lowest available excitation energy, well separated from the continuum spectrum at higher energy.  I  interpret this peak as the signature of the integral charge quasiparticle  recently predicted by Peterson and Jain\cite{Jain05}, and calculate its spectral weight for different filling factors.  
\end{abstract} 
 \maketitle
The fate of one-particle excitations in strongly correlated electronic systems is one of the most basic problems of many-body theory.  Normal electron liquids  in three- and two-dimensional metals and semiconductors support one-particle excitations (quasiparticles)  in spite of strong correlations.  A one-dimensional electron liquid, on the other hand,   does not support such  quasiparticles.\cite{GV05}  

The  two-dimensional electron liquid  (2DEL) at a magnetic field so high that all the electrons reside in the lowest Landau level (LLL)  (filling factor $\nu = \frac{N}{N_L} <1$, where $N_L$ is the degeneracy of the Landau level) is a perfect example of a system in which electron correlations lead to qualitatively new features.\cite{QuantumHall}   At  filling factors of the form $\nu = \frac{n}{2np+1}$, with $n$ and $p$  integers,  the electrons form an incompressible quantum liquid with fractionally charged excitations\cite{Tsui82,Laughlin83} which obey unconventional statistics.\cite{Arovas84}   At the same time the neutral excitation spectrum exhibits a gap.\cite{Girvin86}  This and other  striking properties of   the fractional quantum Hall states are now well understood  in terms of the composite fermion (CF) model.\cite{Jain89,Heinonen98}   

The one-electron properties of the 2DEL at high magnetic field have been  intensively studied, both experimentally, through measurements of the tunneling density of states \cite{Eisenstein92,Ashoori93},  and theoretically, through analytical\cite{Johansson93,He93} and numerical calculations\cite{Hatsugai93,He93,Pikus93}   of the local spectral function.   Neither the experiments nor the early calculations  showed any evidence of the sharp peak in the tunneling density of states that would indicate the existence of a localized quasiparticle excitation of charge $e$.  In a slightly later paper, Haussman\cite{Haussmann96} did find such a peak, but dismissed it as an artifact of the independent boson model on which his calculations (as well as the calculations of Ref. \cite{Johansson93} and ours below) are based.    It was tacitly assumed that  integral-charge quasiparticles must be absent in the fractional quantum Hall liquid.  

Yet, the composite fermion theory of the incompressible quantum Hall liquid suggests a different picture in which single particle excitations of integral charge $e$ can and do exist as bound states of composite fermions.\cite{Jain05}  Consider for example the incompressible liquid at filling factor $\nu=1/3$.  In composite fermion theory this corresponds to one full Landau level of  composite fermions with two vortices attached to each: the  CF filling factor  is  thus $\nu^*=1$.  Let us remove {\it three} CFs  from the fully occupied Landau level: it can be shown\cite{Rezayi87} that the resulting  CF wave function is just what one gets by applying the electron destruction operator to the  ground-state.  We will refer to the state obtained in this manner as the ``single-hole state".   The perfect coincidence of the single hole state with the state of three CF holes implies that the spectral function for removal of one electron has a single peak at the energy required to remove three CFs.  This is very interesting because it suggests that a charge-$e$ hole can exist  as a bound complex of three CF holes.     

Recently, Jain and Peterson\cite{Jain05} have extended the above analysis to the case of electron quasiparticles at $\nu=1/3$, as well as  electron and hole quasiparticles at $\nu=2/5$ and $3/7$ (corresponding, respectively, to two and three full CF-Landau levels).  By carefully extrapolating the results of their  calculations  to the thermodynamic limit they have found that  electron and  hole quasiparticles exist in all cases,   and can be identified with complex bound states of  CFs in different CF Landau levels.   In each case, they find that the strength $Z$ of the quasi-electron peak  is smaller than the strength of the quasi-hole peak.  Furthermore, the intensity of these peaks  decreases rapidly as the size of the $q=0$  gap in the collective excitation spectrum  decreases.  This trend is consistent with the expectation that there should be no quasiparticles  in the compressible state at filling factor $\nu=1/2$ -- the gapless limit of the Jain's sequence.\cite{He93,Halperin93}  Thus, paradoxically, it is precisely the presence of the quantum Hall gap -- the most un-Fermi-liquid-like feature of the quantum Hall liquid -- that ``protects" the electron and the hole quasiparticles  from total collectivization.  Jain and Peterson\cite{Jain05} have further pointed out that the existence of these quasiparticles could be observed in vertical tunneling experiments,  and discussed the reasons why they may have been missed in the experiments carried out so far.

The purpose of this paper is to rederive the existence of integral charge electron/hole quasiparticles in the LLL by what appears  to be a completely different approach, since it does not make use of the notion of composite fermions.   To this end I present an approximate calculation of the one-electron spectral function in the sector of zero angular momentum.   At the absolute zero of temperature this function splits into the sum of two non-overlapping terms, one of which vanishes for negative frequency and the other for positive frequency.  The positive-frequency part is defined as follows:
\begin{equation}
A_>(\omega) = \sum_{n}\vert \langle n|\hat a^\dagger_0|0\rangle\vert^2 \delta(\omega - \omega_{n0})~,
\end{equation}
where $\hat a^\dagger_0$ is the creation operator for the state of angular momentum $m=0$ (in the circularly symmetric gauge), the sum runs over the exact eigenstates $|n\rangle$  ($|0\rangle$ being the ground-state),  and $\hbar \omega_{n0}$ are excitation energies.  The negative-frequency part ($A_<(\omega)$) is obtained by replacing $\hat a^\dagger_0$ with $\hat a_0$ and $\omega$ with $-\omega$ in the above formula.
I will show that this function contains a $\delta$-function peak at the lowest available excitation energy in the $m=0$ sector, and that this peak is well separated from the rest of the  excitation spectrum - a fact already noticed  by Haussmann\cite{Haussmann96}, and dismissed by him as an artifact of the ``independent boson model".    Since, by construction, the spectral function measures the distribution of charge-$\pm e$ excitations over exact eigenstates of the system,  I believe that  the split-off $\delta$-function peak should be interpreted as the signature  of  a long-lived quasiparticle of charge $\pm e$ - the Peterson-Jain quasiparticle\cite{Jain05}.   Notice that this quasiparticle has more energy than the ``conventional"  fractionally charged quasi-electron or quasi-hole, but it can nevertheless be produced (if sufficient energy is available) with a probability proportional to the height of the peak.  The process is analogous to the production of complex bound states of elementary particles  in high energy physics.

One may ask why the peculiar structure of $A(\omega)$ described above was not noticed in earlier theoretical calculations of the spectral function.  These calculations fall into two classes:  exact diagonalizations of few-electron systems\cite{Hatsugai93,He93} and approximate calculations for the infinite system\cite{Johansson93,He93}.  The former are exact but, by their very nature, produce an output that consists of a discrete set of spectral lines separated by gaps.  It is therefore impossible to notice the presence of a single spectral line separated from a continuum, for the simple reason that there is no continuum.  The analytical calculations, on the other hand, were based on approximate bosonization schemes in which  the dispersion of the ``bosons" was assumed to be gapless or, in one case, diffusive.   
Our calculation (as well as that of Haussmann\cite{Haussmann96}) is also based on an approximate bosonization scheme, but the crucial difference is that  the bosons (which represents the neutral collective density oscillations of the liquid) are gapped.   As we show below, the presence of the gap in the boson spectrum is the single feature that causes the appearance of the split-off peak in the spectral function.

Our starting point is the collective description of the quantum Hall liquid as an elastic medium that supports gapped collective excitation\cite{Conti98}.  The Hamiltonian (projected in the LLL)  has the form
\begin{equation}
\hat H_0= \sum_{\vec q} \hbar \omega_{q} \hat b^\dagger_{\vec q}\hat b_{\vec q}~,
\end{equation}
where   $\omega_ q$ is the  frequency of the collective modes at wave vector  $\vec q$, and we assume the dispersion has a gap, i.e., $\omega_q > \omega_{min}$ for all $\vec q$.    The boson operators $\hat b_{\vec q}$ and $\hat b^\dagger_{\vec q}$ are related to the components of the elastic displacement field operator $\hat {\vec u}(\vec r)$ in the following manner:
\begin{equation} \label{ubosonic}
\hat {\vec u}(\vec r) = {\ell \over \sqrt{n_0 {\cal A}}} \sum_{\vec q} \left 
(\hat b_{\vec q}~\vec u(\vec q) e^{i \vec q \cdot \vec r}+ \hat
b^\dagger_{\vec q}~\vec u^*(\vec q) e^{-i \vec q \cdot \vec r}\right)~.
\end{equation}
Here $n_0$ is the average density,  $\ell = \sqrt{\frac{\hbar c}{eB}}$ is the magnetic length, ${\cal A}$ is the area of the system, and $\vec u(\vec q)$ --  the eigenfunction of the associated classical dynamical problem\cite{GV05} -- is given by
\begin{equation}
\vec u(\vec q)  = -i \vec q \ell \sqrt{\frac{{\cal S}}{2 \hbar \omega_q n_0}}+ \frac {\vec e_z \times \vec q}{q^2 \ell} \sqrt{\frac{\hbar \omega_q n_0}{2 {\cal S}}}~,
\end{equation}
where $\vec e_z$ is the unit vector perpendicular to the plane of the two-dimensional electron gas, and  ${\cal S}$ is the dynamical shear modulus.  Notice that, in order to have a collective dynamics, the  medium must be assigned a finite shear modulus:  this assignment is consistent with the liquid nature of the system as long as we work at finite frequencies,  comparable to gap.  Also notice that the bosonic commutation relation $[\hat b_{\vec q},\hat b^\dagger_{\vec q'}] =\delta_{\vec q\vec q'}$  is simply a way of saying that the $x$ and $y$ components of the projected displacement field are canonically conjugate variables: $[\hat u_x(\vec r), \hat u_y(\vec r')] =    -i \frac{\ell^2}{n_0} \delta(\vec r - \vec r') $.  

In order to investigate the possible existence of electron and hole quasiparticles we consider the behavior of a single ``test electron" added to  the system in the coherent state  of angular momentum $m=0$ in the LLL.  This electron interacts with the density fluctuations of the medium via the Coulomb interaction, and since the density fluctuation is related to the displacement field by  $\delta \hat n(\vec r) =-n_0 \vec \nabla \cdot \hat{\vec u}(\vec r)$  we see that the complete hamiltonian for the test electron plus the electron liquid has the form
\begin{equation}\label{hamiltonian}
\hat H = \sum_{\vec q} \hbar \omega_{q} \hat b^\dagger_{\vec q}\hat b_{\vec q} + \sum_{\vec q} M_{q}\left(\hat b_{\vec q} + \hat b^\dagger_{-\vec q} \right )\hat a^\dagger_0 \hat a_0~,
\end{equation}
where $M_{q}$ is an ``electron-phonon matrix element", which, for Coulomb interaction, has the form
\begin{equation}\label{epmatrixelement}
M_q= -\frac{2 \pi e^2 q \ell^2}{\epsilon_b \sqrt{{\cal A}}}
\sqrt{\frac{{\cal S}}{2 \hbar \omega_q}}e^{-q^2 \ell^2/2}~.
\end{equation}
\begin{figure}\label{SpectralFunctionPlot}
\includegraphics[width=6cm]{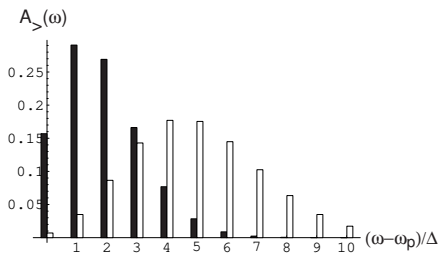}
\caption{The spectral function calculated from Eq.~(\ref{spectralfunction}) at $\nu=1/3$, $\bar \Delta=0.15$ (black bars) and $\nu =2/5$, $\bar \Delta=0.087$ (white bars).} 
\end{figure} 
 It was shown in Refs.~\cite{Conti98,GV05} that, for states described by the Laughlin wavefunction at filling factor $\nu=1/M$, where $M$ is an odd integer, the dynamical shear modulus is related to the $q=0$ gap,  $\Delta = \lim_{q \to 0}\hbar \omega_q$,  by the simple formula 
 \begin{equation} \label{Laughlindelta}
 \frac{{\cal S}}{n_0 \Delta} = \frac  {1-\nu}{4 \nu}~,
 \end{equation}
 For different   filling factors we can either continue to use this relation, or use the shear modulus of the classical Wigner crystal, which is given by $\frac{{\cal S}}{n_0} = 0.0977 \sqrt{\nu} \frac{e^2}{\epsilon_b \ell}$.\cite{Bonsall77}
  
 It should be noted at this point that the hamiltonian $\hat H$ is formally similar to the polaron-problem hamiltonian in which a free electron interacts with optical phonons.  When the polaron hamiltonian is generalized to describe a gas of free electrons interacting with collective density and spin-fluctuations it yields a pseudo-hamiltonian which can be solved (to second order in the ``electron-phonon interaction") leading to the standard theory of the normal Fermi liquid in 3D, with long-lived electron and hole quasiparticles of effective mass $m^*$.\cite{GV05,Asgari05}  The present hamiltonian has the crucial advantage that virtual transitions of the test electron to different single-particle states in the LLL are not allowed:  the electron can never be scattered out of the $m=0$ state.  It is therefore possible to obtain a complete analytic solution of the model Hamiltonian and to calculate the spectral function exactly.
 
The steps of this solution are well known.\cite{GV05} The electron-phonon coupling in the hamiltonian~(\ref{hamiltonian}) can be  eliminated by a unitary transformation.   The new hamiltonian (still written in terms of the original operators) takes the form
 \begin{equation}
 \hat{\bar H} =  \sum_{\vec q} \hbar \omega_{q} \hat b^\dagger_{\vec q}\hat b_{\vec q} +E_p\hat a^\dagger_0\hat a_0~,
 \end{equation}
 where $E_p \equiv -\sum_{\vec q} \frac{M_q^2}{\hbar \omega_q}$ is the polaron shift.  $E_p$ can be calculated analytically if one approximates  $\hbar \omega_q \simeq  \Delta$, by making use of the Laughlin-wave-function-based expression~(\ref{Laughlindelta}) for $\frac{{\cal S}}{n_0\Delta}$. \footnote{Otherwise, using the shear modulus of the Wigner crystal,  we get $E_p =- \frac{0.0244\nu^{3/2}}{\bar \Delta^2}$.} The result is 
\begin{equation}\label{polaronshift}
 E_p =- \frac{1 - \nu}{16 \bar \Delta} \frac{e^2}{\epsilon_b \ell}
 \end{equation} (where $\bar \Delta$ is the gap $\Delta$ expressed in units of $\frac{e^2}{\epsilon_b \ell}$) if one  Notice that $E_p$ is the addition energy of our model, i.e. the difference between the ground-state energies of  the system with and without the ``test electron".   Its value is not expected to be quantitavely accurate since the elastic model is only valid at long wavelengths and certainly misses much of the short-range correlations that control the  value of the ground-state energy.

Let us now proceed to the calculation of the spectral function.   A standard calculation leads to the following result for  $A_>(\omega)$:
 \begin{equation}\label{Agreat}
 A_>(\omega) = \int_{-\infty}^{+\infty}\frac{dt}{2 \pi}e^{i(\omega - \omega_p)t+\sum_{\vec q}\left(\frac{M_q}{\hbar \omega_q}\right)^2 (e^{-i\omega_qt}-1)}~,
 \end{equation}
 where $\omega_p = E_p/\hbar$.  $A_<(\omega)$ is just the mirror image of $A_>(\omega)$ with respect to $\omega = 0$.
 
 A more effective way to write Eq.~(\ref{Agreat}) is
  \begin{equation} \label{spectralfunction.smart}
 A_>(\omega) = Z \delta(\omega-\omega_p) + Z\sum_{k=1}^\infty \frac{g_k(\omega-\omega_p)}{k!}~,
 \end{equation}
 where 
 \begin{equation}\label{renormalizationconstant}
 Z=e^{-\sum_{\vec q} \left( \frac{M_q}{\hbar \omega_q}\right)^2}~,
 \end{equation}
 \begin{equation}\label{defgk1}
 g_1(\omega) = \sum_{\vec q}\left(\frac{M_q}{\hbar \omega_q}\right)^2 \delta(\omega-\omega_q)~,
 \end{equation}
and
\begin{equation}\label{defgk}
g_k(\omega)= \int_0^\infty d \omega'  g_1(\omega-\omega')g_{k-1}(\omega')~, 
\end{equation}
for $k>1$.
From this we see that the spectral function consists of two parts:  a $\delta$-function peak at $\omega = \omega_p$  (first term on the right-hand side of  Eq.~(\ref{spectralfunction.smart}) and a continuum (second term).    The continuum is a sum of terms $g_k(\omega - \omega_p)/k!$  which can be evaluated recursively from Eqs.~(\ref{defgk1}) and (\ref{defgk}).  The $k$-th term of this sum ($k \geq 1$)  differs from zero in a range of frequencies going from $\omega_p+k \omega_{min}$ to $\omega_p+k \omega_{max}$, where $\omega_{min}$ and $\omega_{max}$ are the minimum and maximum values of $\omega_{\vec q}$.   Assuming that the ``bandwidth" of the collective mode, $\omega_{max}-\omega_{min}$ is larger than $\omega_{min}$,  we see that contributions with different $k$ overlap, resulting in a smooth variation of the spectral function for $\omega>\omega_p+\omega_{min}$.  However, the $\delta$-function  at $\omega = \omega_p$ is split-off,  since $\omega_p$ is lower than $\omega_p+\omega_{min}$.  In other words, the presence of an isolated spectral line at $\omega = \omega_p$ is inescapable, and independent of the details of the collective mode dispersion, as long as the latter has a gap . This is the  quasiparticle peak.  Its strength, $Z$,  is the probability of the electron entering the system in the lowest energy state,  with all the magneto-plasmon oscillators remaining in their ground-state.    On the other hand, the area under the $k$-th term of the sum in Eq.~(\ref{spectralfunction.smart}) is the probability of the  electron entering the system in an excited state containing $k$ quanta of the density oscillation field.   

 An analytical evaluation of Eq.~(\ref{spectralfunction.smart}) becomes possible if one neglects the dispersion of the density oscillations, setting $\omega_q = \Delta/\hbar$.   The result is then  
 \begin{equation}\label{spectralfunction}
 A_>(\omega)= Z \sum_{k=0}^\infty \frac{1}{k!} \left(\frac{|E_p|}{\Delta}\right)^k \delta(\omega  - k \Delta/\hbar - \omega_p)~,
 \end{equation}
 where 
 \begin{equation}\label{renormalizationconstant}
 Z= e^{-|E_p|/\Delta}~.
 \end{equation}
 In this approximation, and only in this approximation, the spectral function consists of a series of sharp spectral lines, as shown in Fig. 1.  All these sharp lines are artifacts of the dispersionless approximation, with the notable exception of the zero-phonon line, which, as discussed above, is present even when the dispersion of the collective modes is fully taken into account.     Notice that the excitation probability attains its maximum  when the total number of excitations is as close as possible to $E_p/\Delta$.    As the gap ($\Delta$ in the present approximation)  tends to zero, the injected electron produces, on the average,  larger and larger numbers of collective excitations:  the strength of the zero-magneto-plasmon line tends to zero as the peak of the spectral function shifts to higher and higher excitation levels.   
    
Table I shows  the $q=0$ gap and the renormalization constant calculated from Eqs.~(\ref{polaronshift}) and~(\ref{renormalizationconstant}) for three different values of  $\nu=1/3$ to $2/5$ to $3/7$.  
\begin{table}
\begin{tabular}{|c|c|c|c|}
\hline 
$\nu$& $\bar \Delta$  &  $\bar E_p $& $Z$\\
\hline 
1/3 & $0.15$ &$-0.477$&$0.156$\\
\hline
2/5 & $0.087$ &$-0.431$&$0.007$\\
\hline
3/7 & $0.069$ &$-0.525$&$0.0004$\\
\hline 
\end{tabular}
\caption{\rm{The dimensionless $q=0$ gap $\bar \Delta$, the dimensionless polaron shift $\bar E_p$, and the quasiparticle renormalization constant $Z$ for three different values of the filling factor $\nu$.  $\bar E_p$ is $E_p$ in units of $e^2/\epsilon_b \ell$. The values of $\bar \Delta$ are from Ref.~\cite{Kamilla97}.}}
\label{table1}\end{table}
The rapid decrease of $Z$ with decreasing gap  is consistent with the trend observed by Jain and Peterson in their calculation.\cite{Jain05} However,  in contrast to that calculation, our model does not distinguish between the electron and the hole excitation.  In linear elasticity theory there is complete symmetry between these two types of excitations, since they do not differ in their long-range effect on the elastic medium.  
 
It is remarkable that two very different approaches  --  the composite fermion theory and the continuum elasticity model --  concur in predicting the possibility of charge-$e$ quasiparticles in the LLL.  At a closer inspection, we see some similarities between the two approaches.  In the composite fermion approach, the reduction of $Z$ arises from the fact that there are several excited states of composite fermions that have the same quantum numbers as the electron   (On the other hand, when  one removes one electron at $\nu =1/3$ there is only one  CF bound state with the correct quantum numbers and this is why  $Z=1$ in that case).   The higher up one goes in the number of occupied  CF Landau levels, the larger is the number of CF bound states with the correct quantum numbers:  consequently,  the overlap between the single-electron or single-hole state and and any CF bound state plummets.  Similarly, in the present approach, the reduction of $Z$ is caused by the larger and larger number of magnetophonons that can ``dress up"  the test electron as the gap becomes smaller. 

In conclusion, I have presented qualitative and quantitative arguments supporting the idea that electron/hole quasiparticles of integral charge $e$ should exist in incompressible fractional quantum Hall liquids.  This conclusion is consistent with the recent findings of Jain and Peterson.\cite{Jain05}  The new quasiparticles could  be observed as   conductance resonances in vertical tunneling spectroscopy,\cite{Eisenstein92} as discussed in Ref.~\cite{Jain05}.  

Support from NSF Grant No. DMR-0313681 is gratefully acknowledged.

\end{document}